# Sputtering Current Driven Growth & Transport Characteristics of Superconducting Ti$_{40}$V$_{60}$ Alloy Thin Films

Shekhar Chandra pandey[1,2], Shilpam Sharma[1]*, K. K. Pandey[3], Pooja Gupta[2,4], Sanjay Rai[2,4], Rashmi Singh[5], and M. K. Chattopadhyay [1, 2]

[1]*Free Electron Laser Utilization Laboratory, Raja Ramanna Centre for Advanced Technology, Indore 452013, India.*
[2]*Homi Bhabha National Institute, Training School Complex, Anushaktinagar, Mumbai 400094, India.*
[3]*High Pressure and Synchrotron Radiation Physics Division, Bhabha Atomic Research Centre, Mumbai, 400085, India*
[4]*Indus Synchrotron Utilization Division, Raja Ramanna Centre for Advanced Technology, Indore-452013, India*
[5]*Nano-Functional Materials Lab, Photonic Materials Technology Section, Raja Ramanna Centre for Advanced Technology, Indore 452013, India*

*\* shilpam@rrcat.gov.in*

## Abstract

The room-temperature growth, characterization, and electrical transport properties of magnetron sputtered superconducting Ti$_{40}$V$_{60}$ alloy thin films are presented. The films exhibit low surface roughness and tunable transport properties. As the sputtering current increases, the superconducting transition move towards higher temperatures. Rietveld refinement of two-dimensional XRD (2D XRD) pattern reveals the presence of stress in the films, which shifts from tensile to compressive as the sputtering current increases. Additionally, the crystallite size of the films increases with higher sputtering currents. The films exhibit a strong preferential orientation, contributing to their texturing. The crystallite size and texturing are found to be correlated with the superconducting transition temperature ($T_C$) of the films. As the crystallite size and texturing increase, the $T_C$ of the films also rises.

## 1. Introduction

The binary Ti-V alloys with different compositions exhibit tunable superconducting properties and interesting physical phenomena, which depend on various sample processing parameters. With their proposed utility in the high-field magnets especially for environments having persistent neutron radiation flux[1-3], and the reports of large high-field critical current density ($J_C$) somewhat comparable to that of the existing Nb based superconductors[4,5], the Ti-V alloys are emerging as promising superconducting materials for the engineering applications. The maximum superconducting transition temperature ($T_C$) of these alloys is higher than the $T_C$ of V and Ti and could be explained based on the dynamical electronic correlations in these disordered alloys[6]. As a consequence of the presence of spin fluctuations[7,8], the $T_C$ in these alloys are much smaller than predicted theoretically and the transition widths in these alloys are relatively broad[6]. In their bulk form, these alloys posse excellent mechanical properties,[9] and depending upon the *at%* of Ti, can co-exist in multiple metallurgical phases[6]. For Ti-*at%* greater than 86%, the $\alpha$-phase (hexagonal close packed) is dominant. However for at % up to 68%, the $\beta$-phase (body centred cubic) is the dominant phase. In between 68% to 86%, there is coexistence of $\alpha$– and $\beta$–phases along with additional $\omega$–phase (primitive hexagonal) precipitates[6]. Carefully engineered co-existence of these metallurgical phases aids in enhancing the flux line pinning force density that in turn increases the $J_C$ of these alloys. Enhancement of the $J_C$ of these alloys by intercalation of magnetic and non-magnetic rare-earth elements along with metallurgical treatments,[5,10,11] and successful fabrication of Ti-V alloy based superconducting wires[12] are also reported.

Apart from their utility in the bulk form, careful tuning of the physical properties of Ti-V thin films can make them useful in various technological applications, e.g., thin film getters[13], hydrogen storage[14], RF cavities[15] and superconducting radiation detectors[16]. There are a few reports on the properties of quench condensed Ti-V thin films with $T_C$ of 3.5-4 K[17,18] in their amorphous state. Crystalline films having 40-80% V-content are reported to show superconductivity at temperature ~7 K[17]. Even though the Ti-V alloy thin films deposited and annealed at high temperatures (~700 ºC) show superconductivity at substantially high temperatures of about 12 K[19], the properties of superconducting Ti-V alloy thin films are very scarcely explored[20,21]. Barring these few efforts, to the best of our knowledge, attempts for the optimisation of the properties of Ti-V alloy thin films by varying deposition parameters have not been reported. The tunability of the thin film properties opens the scope of engineering the material for specific applications. In this regard, the room temperature grown Ti-V alloy thin films are expected to be more suitable for their use in various engineering fields and large-scale industries where polycrystalline films are useful, particularly in the case of the substrates that cannot withstand high temperatures. Controlling the average grain sizes, morphology and orientation of the thin films helps in the growth of either homogeneously disordered[22] or granular thin films[23]. The physical properties in the normal and superconducting state like sheet resistance, resistivity, $T_C$, and $J_C$ etc. strongly depend upon the level of disorder in the films and they may be reproducibly controlled by varying the deposition parameters.

Among different physical vapour deposition techniques, magnetron sputtering is a fast and relatively simple thin film growth method that involves quite a few controllable growth parameters to deposit films with the desired properties. Growth parameters such as sputtering current, background argon pressure[23] and negative substrate bias[24] have large impact on the morphology of the thin films. These parameters affect the energy and rate of the incoming ad-atoms. If the growth rate of the films remains lower or comparable to the impurity impingement rate, the film morphology becomes granular with nano-sized grains embedded in an amorphous matrix[23]. The sputtering rate and thus the film growth rate are proportional to the sputtering current. The grain growth is assisted up to a certain magnitude of sputtering current, after which, the grain sizes start to decrease due to supersaturation. Higher sputtering rate leads to the formation of large-sized grains and improvement in the film quality, which could be attained using DC magnetron sputtering process.

In this report, studies on room temperature DC magnetron co-sputtering grown $Ti_{40}V_{60}$ alloy thin films are presented. For optimizing the superconducting and normal state properties of the films, the $Ti_{40}V_{60}$ alloy thin films were deposited using different sputtering currents. By changing the sputtering currents, the granularity and texturing of the films were tuned, as these parameters are expected to have a strong influence on the superconducting and normal state properties of the films. The aim of the present work is to synthesize the optimized thin film having a $T_C$ close to its bulk value, and to improve in the other superconducting parameters of the films. The films were characterized for phase purity, morphology, texturing, DC electrical transport properties. It was observed that adjusting the sputtering current allows the tuning of the superconducting properties of the films. Enhancement in the $T_C$ of the films was noticed with the increase of sputtering current. In addition to the electrical characteristics, sputtering current significantly affects the structural and morphological properties of the films, as the grain orientation and stress within the films is influenced significantly by the sputtering current. Increasing the sputtering current produces better oriented grains, which enhances the texturing in the films.

## 2. Experimental Details

Thin films of $Ti_{40}V_{60}$ alloy were deposited on $SiO_2$ (300 nm, amorphous) coated Si (100) substrates (n-type, 0.5 mm thick) using DC co-sputtering of high purity Ti (99.99 %) and V (99.9 %) targets under ultra-high pure Ar (99.9995 %) residual gas atmosphere. The samples were deposited at ambient temperature. The thin film samples were deposited at different constant DC sputtering currents in a home-built ultra-high vacuum (~ $2\times10^{-8}$ mbar) magnetron sputtering deposition system. The system consists of four confocally arranged 2″ diameter magnetron guns. For ensuring the spatial uniformity of the films, the substrates were mounted on a rotating substrate holder capable of rotating up to 30 rotations per minute (RPM). The system is equipped with a home built load-lock system for the loading and unloading of the samples without disturbing the UHV environment in the main deposition chamber. The 10 mm × 10 mm substrates were ultrasonically cleaned in boiling acetone which was followed by rinsing in ethyl alcohol, rinsing in de-ionized water and finally blow-drying. During deposition, the target to substrate distance was kept constant at 11 cm, the target was rotated at 15 RPM and the background argon pressure was maintained at $2.2\times10^{-3}$ mbar.

To study the changes in the morphology and superconducting properties of alloy thin films as a function of sputtering current, the alloy composition and thickness of the films were kept constant. Since the molar volumes of the individual components in an alloy thin film are proportional to the corresponding thicknesses of those components, the alloy stoichiometry can be estimated from the thickness ratio of the alloy components[25]. Five different samples were deposited by varying the vanadium sputtering current ($I^V$) from 500 mA to 699 mA in the steps of grossly 50 mA. The 500-699 mA range for V sputtering current was selected for the study because we had observed during the optimization process that pure V begins to exhibit metallic properties above 500 mA, while the alloy's $T_C$ starts to degrade above 699 mA. In order to maintain the atomic stoichiometry of the $Ti_{40}V_{60}$ alloy, the sputtering current for the titanium target ($I^{Ti}$) was kept in the range of 352 – 491 mA. The $I^{Ti}$ was estimated using pre-calibrated Ti and V deposition rates (in Å/mA-sec), such that the Ti-content was maintained at 40-*at%*. This pre-calibration was done by measuring the thickness of five thin film samples of Ti and V each, deposited with different sputtering currents keeping the deposition time fixed. The thickness of the films was kept constant at 40 nm by adjusting the time ($t_d$) of Ti and V co-deposition such that for higher sputtering currents, the deposition time is reduced with respect to the total deposition rates of Ti and V. The thicknesses of the films were measured using X-ray reflectivity measurements (XRR) performed using a Bruker, GmbH make D8 diffractometer. The XRR pattern was fitted using REFLEX software[26]. The elemental analysis for confirming the alloy composition was performed using the energy dispersive analysis of x-ray (EDAX) setup attached to the scanning electron microscope (SEM, Carl Zeiss, Germany).

Initially, the phase purity of the alloy films was confirmed using grazing incidence x-ray diffraction (GIXRD) measurements using Bruker, GmbH make D8 diffractometer. Synchrotron based 2D XRD measurements on the thin films were performed using ~ 15 keV X-rays at the Extreme Conditions Angle Dispersive/Energy Dispersive X-ray Diffraction (EC-AD/ED-XRD) beam-line (BL-11) at the Indus-2 synchrotron source, Raja Ramanna Centre for Advanced Technology (RRCAT), Indore, India. The desired wavelength (0.7003 Å) for the 2D XRD experiments was selected from the white light coming from the bending magnet using a Si (111) channel cut monochromator. The monochromatic beam was then focused on to the sample with a Kirkpatrick-Baez (K-B) mirror. A MAR345 image plate detector (which is an area detector) was used to collect two dimensional diffraction data. The sample to detector distance, beam centre, detector tilt and the wavelength of the beam were calibrated using NIST standards $LaB_6$ and $CeO_2$. Calibration and conversion/integration of the 2-D diffraction patterns to the 1-D intensity vs. 2θ plots was done using the FIT2D software[27]. The grain size, stress, strain and texture analysis of the alloy thin films were performed using Rietveld refinement method as implemented in the Materials Analysis Using Diffraction (MAUD) software package[28].

Measurements of electrical resistance as a function of temperature ($R(T)$) was performed in the range 2-300 K using a Cryo-free Spectromag CFSM7T-1.5 magneto-optical cryostat system (Oxford Instruments, UK). The *R(T)* measurements were performed in the Van der Pauw four-probe geometry, where the electrical contacts with thin copper wire probes were made with the help of room temperature curable silver paste.

3. **Results and Discussion**

Previously, we studied the room temperature growth of different compositions of Ti-V alloy thin films having Ti *at*-% 30, 36, 40, 43, and 50%. We found that the composition $Ti_{40}V_{60}$ exhibited the highest transition temperature[29]. This alloy composition is also reported to have the maximum critical current density among all other Ti-V compositions[4]. This composition was chosen to study the effects of different sputtering parameters on the superconducting properties of the Ti-V thin film system. A series of five $Ti_{40}V_{60}$ alloy thin film samples were deposited by varying the sputtering currents for vanadium and titanium as stated in the previous section. The sputtering currents for vanadium and titanium in these five samples are given in table1.

GIXRD measurements were performed to characterise the phase purity of the alloy films. The plots of normalized intensity as a function of diffraction angle are presented in figure 1. All the observed reflections were indexed to a *bcc* crystal structure with space group $Im\bar{3}m$. Notably, individual peaks corresponding to Ti and V were not observed in the GIXRD spectra, suggesting a homogeneous composition within the films. From figure 1, it is also observed that the intensity of the main peak (110) increases marginally with increasing sputtering current. As can be seen from the figure 1, Ti-addition shifts the peaks towards lower $2\theta$ values as compared to the pure vanadium. However despite having the same nominal composition, the peak for different samples are not occurring at the same $2\theta$ value. Typically, changes in the peak intensity are due to variations in the grain size and texture, while shifts in peak positions are indicative of stress within the film. Thus, the observed peak shifts with increasing sputtering current may be attributed to the stress within the film. The stress can be either compressive or tensile, resulting in the lattice parameters being smaller or larger than that of the stress-free sample. This causes corresponding changes in peak positions in accordance with Bragg's law.

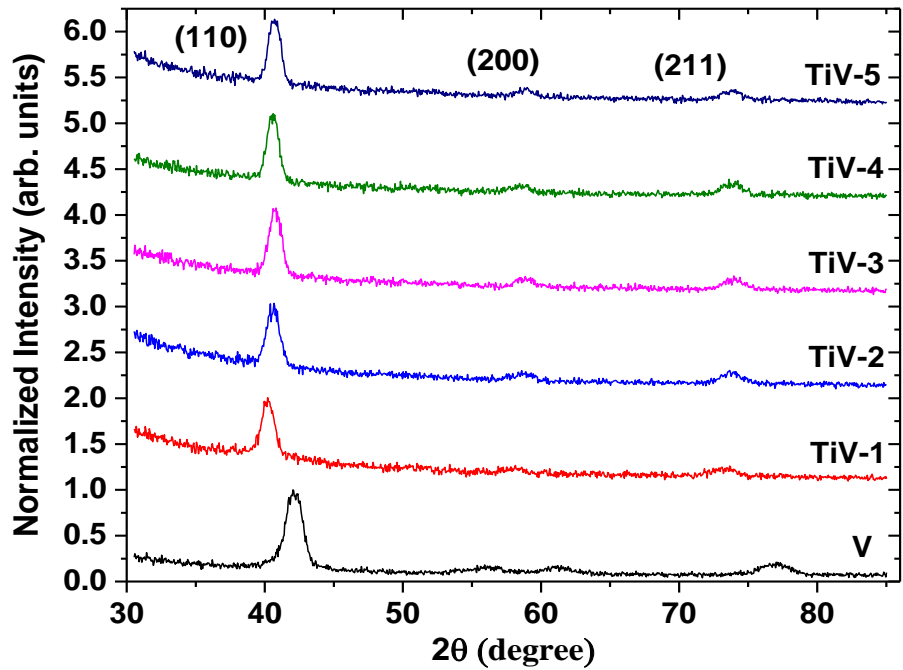

**Figure 1:** GIXRD patterns of Ti-V alloy thin films showing the polycrystalline nature of the films. The XRD peaks of $Ti_{40}V_{60}$ thin films shift to lower $2\theta$ values as compared to pure vanadium film. The (110) peak positions change non-monotonically. The intensity of (110) and (211) peaks increases with the increasing deposition current.

Since the $I^{Ti}$ and $t_d$ for the samples were estimated such that the alloy stoichiometry and film thickness remain constant. Thus, the measured thickness of the samples provide a confirmation for the stoichiometry. The thickness and roughness of the films were precisely determined through the analysis of XRR (x-ray reflectivity) measurements, using the REFLEX software for pattern fitting[26]. Figure 2 presents the fitted XRR patterns, for two different samples. All the films were found to have a thickness of 40 nm and a roughness of 2 nm. This confirms that the five different thin film samples have identical Ti content.

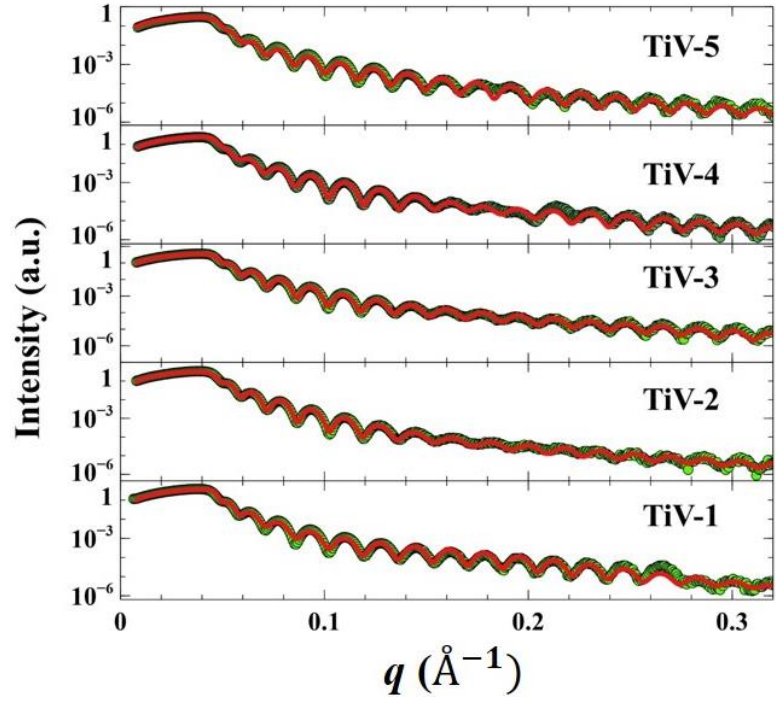

**Figure 2:** The XRR patterns as a function of the scattering vector for the five $Ti_{40}V_{60}$ thin film samples. The thickness and roughness are estimated to be 40 nm and 2 nm respectively through fitting the experimental data. The green curves represent the observed experimental data, while the red curves correspond to the fitted patterns.

To actually measure the Ti content in the samples, high resolution SEM and EDAX measurements were performed. Signature of grain boundaries were not observed in the high-resolution SEM images, indicating that the grain sizes are smaller than the resolution limit of the SEM. Elemental and compositional quantification of the alloy films confirm that the estimated compositions closely match the nominal values. Table 1 shows the vanadium and titanium atomic percentage in the Ti-V alloy thin films.

**Table 1**: Sputtering currents for V and Ti used in the deposition of five $Ti_{40}V_{60}$ alloy thin film samples, along with the individual Ti and V atomic percentages estimated from EDAX measurements.

| Sample No. | Sputtering current | Ti at % | V at% |
|---|---|---|---|
| TiV-1 | $I^V : I^{Ti}$ = 501:352 mA | 38.9 | 61.04 |
| TiV-2 | $I^V : I^{Ti}$ = 557:391 mA | 36.13 | 63.86 |
| TiV-3 | $I^V : I^{Ti}$ = 601:422 mA | 36.9 | 63.07 |
| TiV-4 | $I^V : I^{Ti}$ = 654:459 mA | 38.8 | 61.1 |
| TiV-5 | $I^V : I^{Ti}$ = 699:491 mA | 39.94 | 60.05 |

After confirming that all the samples have similar alloy compositions, the reason for the shifts in the GIXRD peak positions shown in figure 1 could be attributed to the stress in the films. To confirm this experimentally, radial x-ray diffraction (2D XRD) measurements were performed at the BL-11 beam line of Indus-2 synchrotron source at RRCAT. In this technique, an area detector is employed to capture the diffraction spectrum, providing a two-dimensional (2D) image of the diffraction pattern. Figure 3 illustrates one such 2D diffraction pattern obtained for a $Ti_{40}V_{60}$ alloy thin film, representing the intensity distribution of scattered X-rays as a function of Bragg ($2\theta$) and azimuthal ($\chi$) angles (where $\chi$ is the angle along the diffraction ring).

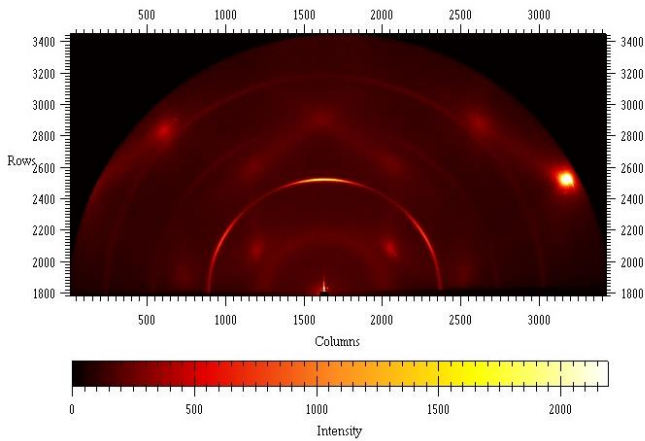

**Figure 3:** Radial XRD image of $Ti_{40}V_{60}$ thin film. Three distinct rings correspond to the three peaks observed from the 2D-XRD measurement.

Three distinct rings are visible in the diffraction pattern corresponding to the three indexed peaks in the GIXRD measurements (refer figure 1). The 2D XRD results were analysed using Rietveld refinement technique. Prior to refinement using MAUD software, the diffraction images were processed using Fit2D and Fit2D2maud software. Utilizing the cake function, we defined the start and end azimuth, and the $2\theta$ ranges for the analysis. For this analysis, the 0-to-180-degree coverage ring was integrated over 5-degree increments of the azimuth angle, resulting in 36 slices of the whole 2D XRD pattern. Analysis of the radial XRD data provides information on the lattice parameter, crystallite sizes, stress values, micro-strain and texturing of the films. For the estimation of the stress, the Triaxial Stress Isotropic E model of the MAUD software was utilised. We restricted stress tensor component $\sigma_{33} = -2\sigma_{11} = -2\sigma_{22}$ and $\sigma_{ij}= 0$ for $i \neq j$. $\sigma_{33}$ is stress component along thin film normal direction and $\sigma_{11}$ and $\sigma_{22}$ are the orthogonal stress components in the plane of the thin film. The model converts the lattice strains to the stress values in the sample using isotropic Young's modulus and Poisson's ratio. The values for Young's modulus and Poison's ratio for the refinement are taken from the literature[30].

The 2D spectra shown in figure 4 displays variation in the peak position with the azimuthal angle, clearly indicating the presence of stress within the film. The 2D plot of the experimentally observed intensity and the calculated intensity after Rietveld refinement for the five $Ti_{40}V_{60}$ samples under study are shown in this figure. The lower portions of the panels represent the experimentally measured spectrum, while the upper portions show the Rietveld refined spectrum. Table 2 shows the values of the different parameters found after the analysis. The TiV-1 sample has a higher lattice parameter value than the other TiV samples. This increased lattice parameter suggests the presence of tensile stress in the film, which will be further validated in the following section.

**Table 2**: Parameters obtained from the Rietveld refinement of the five $Ti_{40}V_{60}$ alloy thin film samples.

| Sample name | Lattice parameter (Å) | | Crystallite size (Å) | | Micostrain | | $\sigma_{33}$ (GPa) | |
|---|---|---|---|---|---|---|---|---|
| | Value | Error | Value | Error | Value | Error | Value | Error |
| TiV-1 | 3.1277 | 1.21E-4 | 197.55 | 1.26 | 0.0033 | 1.9E-5 | 0.2053 | 5.5E-3 |
| TiV-2 | 3.1139 | 0.79E-4 | 233.14 | 1.41 | 0.0041 | 2.4E-5 | -0.1990 | 3.2 E-3 |
| TiV-3 | 3.1171 | 0.75E-4 | 241.15 | 2.09 | 0.0051 | 4.7E-5 | -0.0645 | 3.3 E-3 |
| TiV-4 | 3.1110 | 0.8E-4 | 256.84 | 2.36 | 0.0046 | 4.9E-5 | -0.1961 | 3.3 E-3 |
| TiV-5 | 3.1179 | 0.8E-4 | 268.51 | 1.95 | 0.0047 | 3.6E-5 | -0.1647 | 2.8 E-3 |

As can be observed from figure 4, the film deposited with the lowest sputtering current exhibits a curvature of various peak positions (2$\theta$) to the left side, whereas as the sputtering current increases, the curvature shifts towards the right side. This indicates that the film deposited at the lowest current has tensile stress, whereas, the stress changes to compressive as the deposition current increases. The values of $\sigma_{33}$ given in table 2 also depict the same phenomena. Negative values of $\sigma_{33}$ indicate compressive stress, while positive values indicate tensile stress along thin film normal. Thus, the film deposited with the lowest sputtering current exhibits tensile stress, while as the sputtering current increases, the stress becomes compressive type.

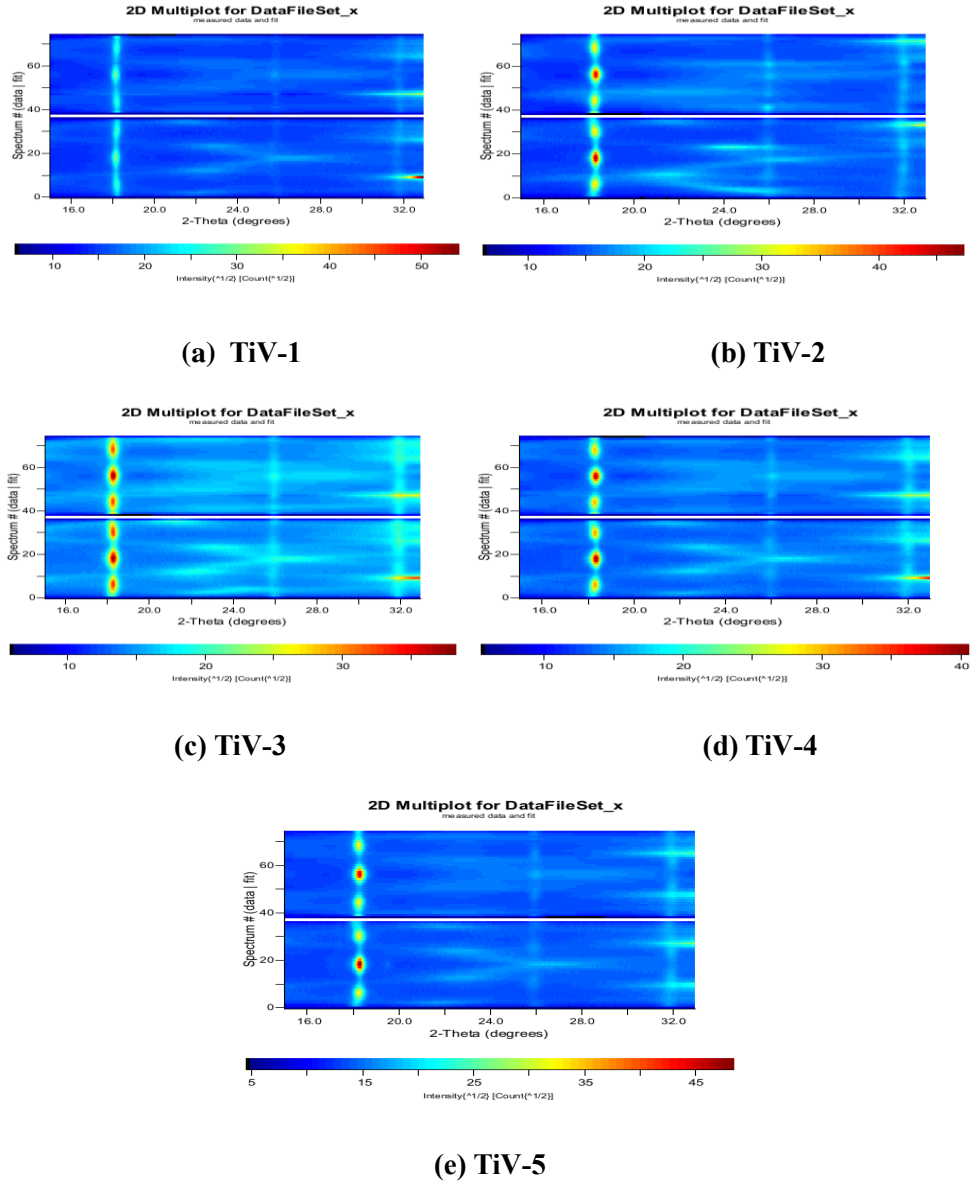

(a) TiV-1

(b) TiV-2

(c) TiV-3

(d) TiV-4

(e) TiV-5

**Figure 4**: 2D plot of the experimentally observed and Rietveld refined intensity for five $Ti_{40}V_{60}$ thin film samples. The lower portions of the panel represents the experimentally measured spectrum, while the upper portions represents the Rietveld refined spectrum.

Polycrystalline materials consist of numerous crystalline grains oriented in various directions, collectively contributing to the overall XRD pattern. The orientation of each grain relative to a reference axis can vary, leading to an orientation distribution that describes how the grains or crystallites are distributed throughout the material. The deviation of the crystallite orientation distribution from that of an ideal single crystal is measured as texture or preferred orientation. A quantitative measure of texture can be presented by plotting the inverse pole figures, which reveal the deviation of the crystallites from a random distribution. This deviation is often quantified in terms of multiple of random distribution (MRD) values. A MRD value of 1 indicates a fully random distribution of the crystallites. Any deviation from 1 indicates texturing in the sample. For texture analysis, we employed the E-WIMV model in MAUD, a tomographic algorithm allowing for incomplete pole figure coverage. Figure 5 shows the plots of inverse pole figures for five Ti-V alloy thin film samples. In general, for a *bcc* crystal, the presence of three corners in the inverse pole figures indicates texturing in the 001 (bottom left), 110 (bottom right), and 111 (top corner) directions. Therefore, the dark red portion in the right corner confirms the strong 110 pole in the normal direction. The sample plane is defined as the RD-TD plane, while the normal to the sample plane is denoted as ND. Here RD, TD, and ND represent the rolling, transverse, and normal directions, respectively.

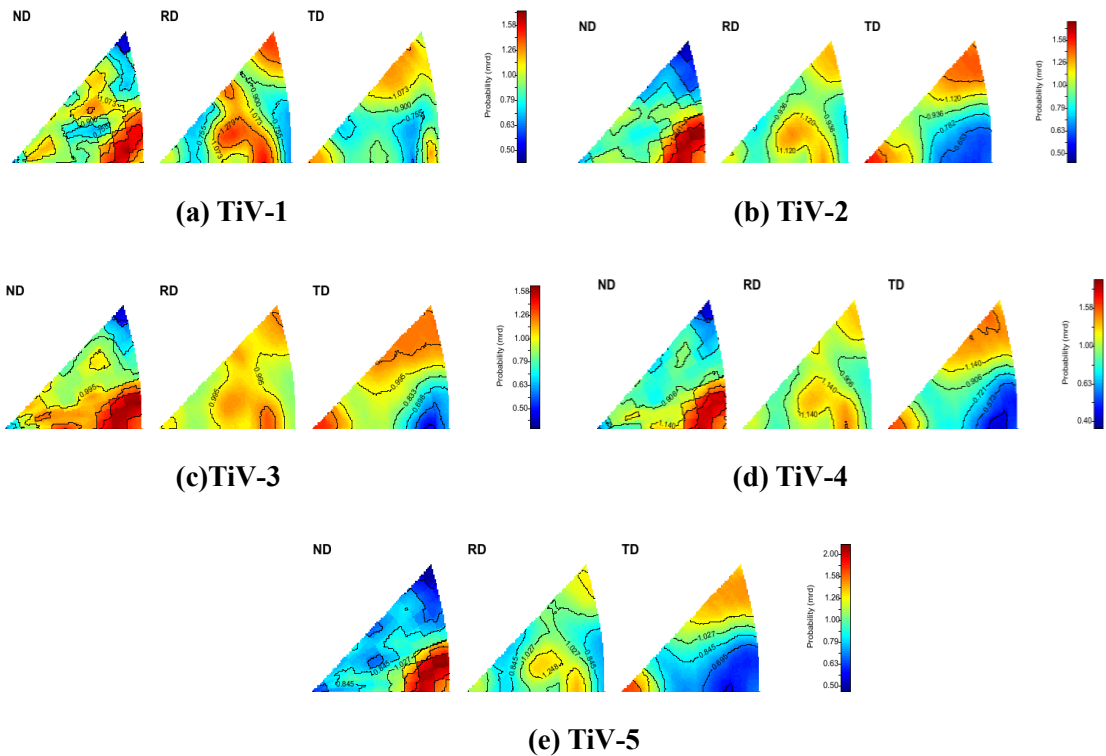

**Figure 5:** Inverse pole figures of the five $Ti_{40}V_{60}$ alloy thin films. The dark red portion in the right corner of the figures confirm the strong 110 pole in the normal direction.

Figure 6(a) illustrates that as the sputtering current increases, there is a corresponding variation in the lattice parameter values. This variation in lattice parameter is shows strong correlation with the stress values shown in Figure 6(b). In the $Ti_{40}V_{60}$ alloys, the reported lattice parameter is 3.15 Å [4]. Under tensile stress, the lattice parameter approaches the reported value, which is seen for the sample with the lowest sputtering current. Under compressive stress, there is a significant deviation from the standard value. Figure 6(c) shows that the crystallite size increases systematically with the sputtering current. Figures 6(d) and 6(e) demonstrate that as the sputtering current increases, the micro-strain and MRD maximum values also increase. This increment in the MRD maximum values with increasing sputtering current indicates a corresponding increase in texturing within the film.

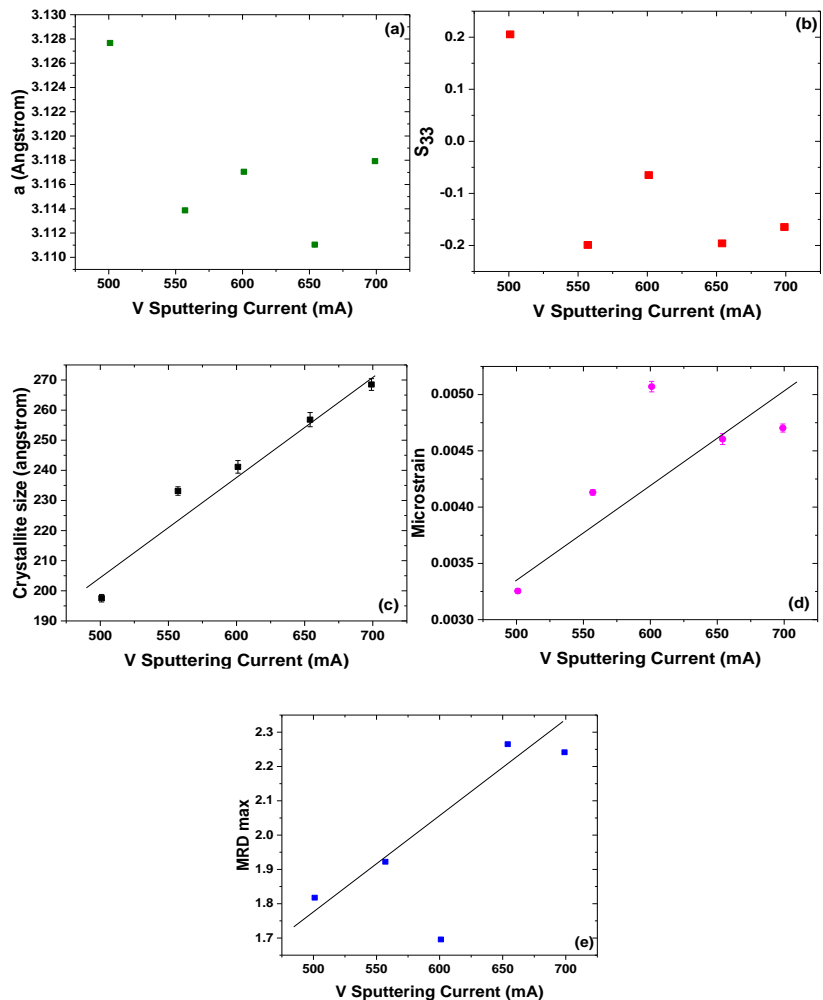

**Figure 6:** Variation of parameters obtained from Rietveld refinement of $Ti_{40}V_{60}$ alloy thin films as a function of vanadium sputtering current. Straight lines in the above figures are guide to the eyes.

The *R(T)* curves for the Ti-V alloy thin films in the temperature range 2- 300 K shown in figure 7

confirm the superconducting nature of the films. It can be seen that all the five $Ti_{40}V_{60}$ alloy thin films show a superconducting transition to the zero resistance state. The highest $T_C \sim 5.3$ K was obtained for the film deposited at highest $I^V$ and $I^{Ti}$. Table 3 shows the variation of $T_C$ with the sputtering currents. As the sputtering current increases, the $T_C$ increases gradually. Notably, the film with tensile stress has the lowest $T_C$. As the stress changes from tensile to compressive, the $T_C$ increases gradually towards the bulk value. Here, we corroborate that the $T_C$ enhancement correlates with the growth of crystallite sizes as sputtering current is increased. Correlating with the MRD values it is observed that the highest $T_C$ also corresponds with the best texturing within the $Ti_{40}V_{60}$ film.

**Table 3:** Superconducting transition temperature of five $Ti_{40}V_{60}$ alloy thin films

| S.No. | Sample name | Sputtering current | $T_C$ (K) |
|---|---|---|---|
| 1. | TiV-1 | $I^V : I^{Ti}$ = 501:352 mA | 2.6 |
| 2. | TiV-2 | $I^V : I^{Ti}$ = 557:391 mA | 4.7 |
| 3. | TiV-3 | $I^V : I^{Ti}$ = 601:422 mA | 4.8 |
| 4. | TiV-4 | $I^V : I^{Ti}$ = 654:459 mA | 5.2 |
| 5. | TiV-5 | $I^V : I^{Ti}$ = 699:491 mA | 5.3 |

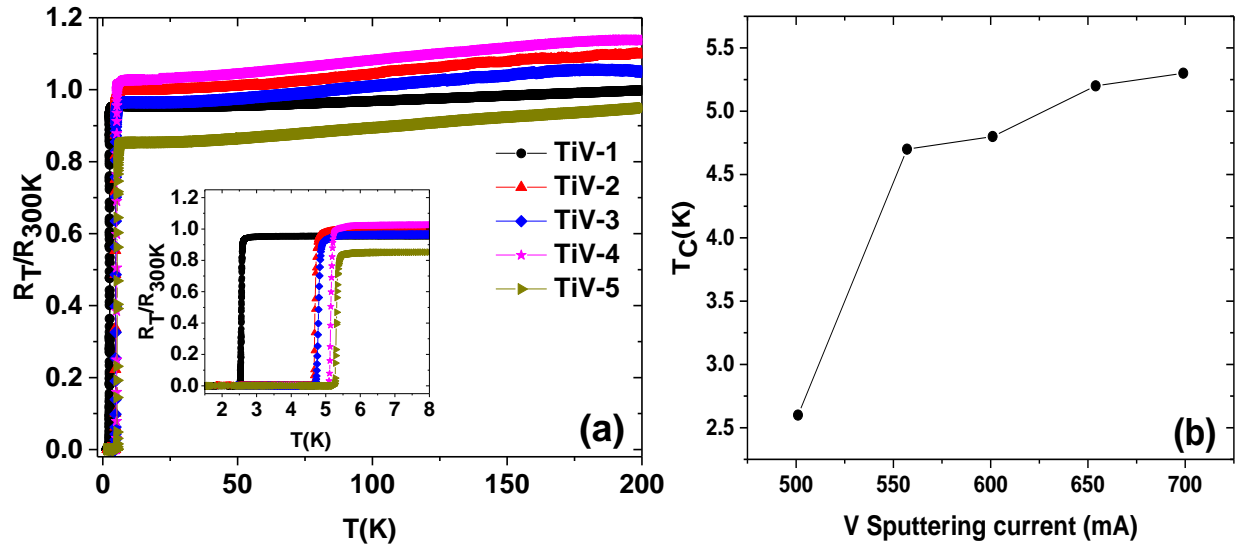

**Figure 7:** (a) The variation in resistance as a function of temperature ($R(T)$) for $Ti_{40}V_{60}$ alloy thin films is shown in the temperature range of 2-200 K. Inset: $R(T)$ near the $T_C$, highlighting that all $Ti_{40}V_{60}$ alloy thin films exhibit a superconducting transition to the zero-resistance state. (b) The $T_C$ of the films increases with increasing vanadium sputtering current.

## 4. Conclusion

The Ti$_{40}$V$_{60}$ alloy thin films with controlled tuning of transition temperature, grain morphology and texturing have been successfully deposited at ambient temperatures. The samples were deposited with very low surface roughness of around 2 nm and good adhesion. The $T_C$ of the samples was found to increase from 2.6 K to 5.3 K with the increase in the sputtering current. This increase in the $T_C$ of the samples could be attributed to the increase in the grain size and texturing of the thin films. A change from tensile to compressive stress was observed as the sputtering current increases. This change in the type of stress could be utilized for tuning the mechanical and transport properties of superconducting thin films with desired composition.

## Authors' Contributions

All authors contributed equally to this work.

## Acknowledgement

We would like to thank Dr. V. Srihari for the 2D XRD measurement and assistance with data analysis, Dr. V. R. Reddy for the XRR measurement, and Dr. L. S. Sarath Chandra for his valuable discussions.

## Data Availability

The data that support the findings of this study are available from the corresponding author upon reasonable request.

## References

[1] M. Tai, K. Inoue, A. Kikuchi, T. Takeuchi, T. Kiyoshi, and Y. Hishinuma, IEEE Transactions on Applied Superconductivity **17,** 2542 (2007).
[2] L. I. Ivanov, V. V. Ivanov, V. M. Lazorenko, Y. M. Platov, and V. I. Tovtin, Journal of Nuclear Materials **191,** 928 (1992).
[3] Y. Higashiguchi, H. Kayano, and S. Morozumi, Journal of Nuclear Materials **133,** 662 (1985).
[4] M. Matin, L. S. Sharath Chandra, M. K. Chattopadhyay, R. K. Meena, R. Kaul, M. N. Singh, A. K. Sinha, and S. B. Roy, Physica C: Superconductivity and its Applications **512,** 32 (2015).
[5] S. Ramjan, L. S. Sharath Chandra, R. Singh, P. Ganesh, A. Sagdeo, and M. K. Chattopadhyay, Journal of Applied Physics **131,** 063901 (2022).
[6] D. Jones, A. Östlin, A. Weh, F. Beiuşeanu, U. Eckern, L. Vitos, and L. Chioncel, Physical Review B **109,** 165107 (2024).
[7] M. Matin, L. S. Sharath Chandra, S. K. Pandey, M. K. Chattopadhyay, and S. B. Roy, The European Physical Journal B **87,** 131 (2014).
[8] M. Matin, L. S. Sharath Chandra, R. Meena, M. K. Chattopadhyay, A. K. Sinha, M. N. Singh, and S. B. Roy, Physica B: Condensed Matter **436,** 20 (2014).
[9] R. E. Gold and R. Bajaj, Journal of Nuclear Materials **122,** 759 (1984).

10. S. Paul, S. Ramjan, R. Venkatesh, L. S. S. Chandra, and M. K. Chattopadhyay, IEEE Transactions on Applied Superconductivity **31,** 1 (2021).
11. S. K. Ramjan, A. Khandelwal, S. Paul, L. S. S. Chandra, R. Singh, R. Venkatesh, K. Kumar, R. Rawat, S. Dutt, A. Sagdeo, P. Ganesh, and M. K. Chattopadhyay, Journal of Alloys and Compounds **976,** 173321 (2024).
12. T. Takeuchi, H. Takigawa, M. Nakagawa, N. Banno, K. Inoue, Y. Iijima, and A. Kikuchi, Superconductor Science and Technology **21,** 025004 (2008).
13. A. Boyko, D. Gaev, S. Timoshenkov, Y. Chaplygin, and V. Petrov, Materials Sciences and Applications **04,** 57 (2013).
14. K. Edalati, H. Shao, H. Emami, H. Iwaoka, E. Akiba, and Z. Horita, International Journal of Hydrogen Energy **41,** 8917 (2016).
15. V. Palmieri, in *New materials for superconducting radiofrequency cavities*, 2001, p. 162.
16. H. Geoffray, A. Monfardini, S. Marnieros, M. Piat, L. Rodriguez, and A. Bardoux, *CNES detector developments from far-infrared to mm: status and roadmap*, Vol. 9153 (SPIE, 2014).
17. B. Bandyopadhyay, P. Watson, Y. Bo, and D. G. Naugle, Journal of Physics F: Metal Physics **17,** 433 (1987).
18. P. Watson, B. Bandyopadhyay, Y. Bo, D. Rathnayaka, and D. G. Naugle, Materials Science and Engineering **99,** 175 (1988).
19. H. J. Spitzer, in *Low Temperature Physics-LT 13*, edited by K. D. Timmerhaus, W. J. O'Sullivan, and E. F. Hammel (Springer US, Boston, MA, 1974), p. 485.
20. P. M. Tedrow and R. Meservey, Physics Letters A **69,** 285 (1978).
21. J. Edgecumbe, L. G. Rosner, and D. E. Anderson, Journal of Applied Physics **35,** 2198 (1964).
22. M. Chand, G. Saraswat, A. Kamlapure, M. Mondal, S. Kumar, J. Jesudasan, V. Bagwe, L. Benfatto, V. Tripathi, and P. Raychaudhuri, Physical Review B **85,** 014508 (2012).
23. S. Sharma, E. P. Amaladass, N. Sharma, V. Harimohan, S. Amirthapandian, and A. Mani, Physica B: Condensed Matter **514,** 89 (2017).
24. S. Sharma, S. Abhirami, E. P. Amaladass, and A. Mani, Materials Research Express **5,** 096401 (2018).
25. W. Zhang, A. T. Bollinger, R. Li, K. Kisslinger, X. Tong, M. Liu, and C. T. Black, Scientific Reports **11,** 2358 (2021).
26. G. Vignaud and A. Gibaud, Journal of Applied Crystallography **52,** 201 (2019).
27. A. P. Hammersley, S. O. Svensson, M. Hanfland, A. N. Fitch, and D. Hausermann, High Pressure Research **14,** 235 (1996).
28. L. Lutterotti, Nuclear Instruments and Methods in Physics Research Section B: Beam Interactions with Materials and Atoms **268,** 334 (2010).
29. S. C. Pandey, S. Sharma, A. Khandelwal, and M. K. Chattopadhyay, arXiv [cond-mat.supr-con] (2023).
30. P. Majumdar, S. B. Singh, and M. Chakraborty, Materials Science and Engineering: A **489,** 419 (2008).